\documentclass[final,5p,times,twocolumn]{elsarticle}

\usepackage[utf8]{inputenc}
\usepackage[numbers]{natbib}
\usepackage{pdflscape}

\usepackage{float}

\usepackage{graphicx}
\usepackage{tabularx}
\usepackage{adjustbox}
\usepackage{tikz}
\usepackage{forest}
\usepackage{titlesec}
\usepackage{longtable}
\usepackage{makecell}
\usepackage{amssymb}
\usepackage{amsmath} 
\usepackage{multirow} 
\usepackage{xcolor}
\usepackage{enumitem}
\usepackage{svg}
\usepackage{booktabs} 
\usepackage[colorlinks=true, linkcolor=black, citecolor=black, urlcolor=black]{hyperref}

\usepackage{bm}
\usepackage{stmaryrd}
\DeclareMathOperator*{\att}{att}
\DeclareMathOperator*{\TE}{TE}

\DeclareMathOperator*{\concat}{concat}
\DeclareMathOperator*{\disc}{DISC}
\DeclareMathOperator*{\coral}{CORAL}
\DeclareMathOperator*{\mmd}{MMD}
\DeclareMathOperator*{\mse}{MSE}
\DeclareMathOperator*{\seg}{SEG}

\definecolor{darkgreen}{RGB}{0, 100, 0}  
\definecolor{lightviolet}{RGB}{230, 230, 250} 
\definecolor{violet}{RGB}{199, 184, 230} 
\definecolor{darkviolet}{RGB}{226,183,226} 

\usepackage{color}
\titleformat*{\section}{\normalsize\bfseries}
\titleclass{\subsubsubsection}{straight}[\subsection]
\newcounter{subsubsubsection}[subsubsection]
\renewcommand{\thesubsubsubsection}{\thesubsubsection.\arabic{subsubsubsection}}
\titleformat{\subsubsubsection}
 {\normalfont\normalsize\itshape}{\thesubsubsubsection}{1em}{}
\titlespacing*{\subsubsubsection}
{0pt}{3.25ex plus 1ex minus .2ex}{1.5ex plus .2ex}

\usetikzlibrary{shapes,arrows.meta,positioning}
\tikzstyle{block} = [rectangle, draw, fill=blue!20, text width=5em, text centered, rounded corners, minimum height=4em]
\tikzstyle{line} = [draw, -Latex]

\date{}
\journal{arXiv}

\begin{document}
\let\WriteBookmarks\relax
\def\floatpagepagefraction{1}
\def\textpagefraction{.001}

\begin{frontmatter}
\title{Deep learning-enabled prediction of surgical errors during cataract surgery: from simulation to real-world application}

\author[1,2]{Maxime Faure}
\ead{maxime.faure@univ-brest.fr}
\author[1,3]{Pierre-Henri Conze}
\author[1,2,4]{Béatrice Cochener}
\author[1,2,4]{Anas-Alexis Benyoussef}
\author[1,2]{Mathieu Lamard}
\author[1]{Gwenolé Quellec}

\address[1]{LaTIM UMR 1101, Inserm, Brest, France}
\address[2]{University of Western Brittany, Brest, France}
\address[3]{IMT Atlantique, Brest, France}
\address[4]{Ophtalmology Department, University Hospital of Brest, Brest, France}

\begin{abstract}
Real-time prediction of technical errors from cataract surgical videos can be highly beneficial, particularly for telementoring, which involves remote guidance and mentoring through digital platforms. However, the rarity of surgical errors makes their detection and analysis challenging using artificial intelligence. To tackle this issue, we leveraged videos from the EyeSi Surgical cataract surgery simulator to learn to predict errors and transfer the acquired knowledge to real-world surgical contexts. By employing deep learning models, we demonstrated the feasibility of making real-time predictions using simulator data with a very short temporal history, enabling on-the-fly computations. We then transferred these insights to real-world settings through unsupervised domain adaptation, without relying on labeled videos from real surgeries for training, which are limited. This was achieved by aligning video clips from the simulator with real-world footage and pre-training the models using pretext tasks on both simulated and real surgical data. For a 1-second prediction window on the simulator, we achieved an overall AUC of 0.820 for error prediction using 600$\times$600 pixel images, and 0.784 using smaller 299$\times$299 pixel images. In real-world settings, we obtained an AUC of up to 0.663 with domain adaptation, marking an improvement over direct model application without adaptation, which yielded an AUC of 0.578. To our knowledge, this is the first work to address the tasks of learning surgical error prediction on a simulator using video data only and transferring this knowledge to real-world cataract surgery.
\end{abstract}


\onecolumn


\end{frontmatter}

\section{Introduction}

\subsection{Context}

Cataract, defined as the opacification of the crystalline lens, is a prevalent ocular condition that significantly impacts vision worldwide \citep{Rossi2021}. Cataract surgery is the most performed surgical procedure worldwide, highlighting its critical role in ophthalmology and public health \citep{Eurostat2023}. The large volume of procedures has generated extensive data, offering opportunities to improve quality management, education, and training \citep{SimMuller2023}. In practice, phacoemulsification is the most effective method for lens removal \citep{Davis2016}, involving the replacement of the clouded lens with an artificial one. Successful execution of the capsulorhexis step, a crucial stage in cataract surgery, is critical to prevent complications. This technique involves creating a circular opening in the anterior lens capsule to access and remove the cloudy lens. For optimal results, the capsulorhexis must be circular, regular, and well-centered, without radial extensions or capsular tags which can lead to surgical complications \citep{Gimbel1991}. Advances in artificial intelligence (AI), particularly deep learning, have demonstrated potential in evaluating surgeon performance through video analysis \citep{Rampat2021}. These technologies have been shown to objectively assess surgical skills and support intraoperative tasks such as real-time tool identification \citep{ALHAJJ2018203,Matton2022}, and surgical phase recognition \citep{QUELLEC2014579, Quellec2015, Touma2022}. Video analysis offers benefits beyond the operating room, such as remote surgical supervision and telementoring, and has shown promise in detecting specific errors, including those related to intraocular lens (IOL) implantation \citep{ghamsarian2021lensid}, continuous curvilinear capsulorhexis and phacoemulsification \citep{Morita2020}. However, there remains a lack of comprehensive data on surgical errors. While public datasets like CaDIS \citep{GRAMMATIKOPOULOU2021102053}, 101-Cataract \citep{DBLPconfmmsysSchoeffmannTSMP18}, and Cataract-1K \citep{Ghamsarian2024} exist, the scarcity of annotated data for surgical error detection remains a major challenge. Surgical simulators have emerged as valuable tools to address this limitation and generate substantial data for analysis. These simulators provide a controlled environment for collecting extensive datasets, enabling the study of surgical techniques and error detection. Various simulation methods are employed in cataract surgery training, including virtual reality, wet-lab, dry-lab models, and e-learning for technical and non-technical skills \citep{Lee2020}. Among these, the EyeSi simulator is the most widely used \citep{CISSECECILEANG}. The effectiveness of the EyeSi simulator in ophthalmology training has been widely demonstrated. It accurately replicates many aspects of cataract surgery, providing essential training for residents \citep{CARR2024172}. Simulator-based training has also been shown to reduce operative times for surgical residents learning phacoemulsification, compared to traditional methods \citep{Ahmed2020}. A strong link between simulator performance and real-life outcomes has been established. \cite{Bozkurt_Oflaz2018} found that the number of surgeries performed by surgeons — reflecting their practical experience — correlated with simulator scores. Studies have shown correlations between simulator scores and various metrics, such as GRASIS \citep{Cremers2005} scores \citep{ROOHIPOOR20171105}, motion-tracking performance \citep{Thomsen2017}, and OSACSS \citep{Nuova2020} scores, effectively distinguishing between novice and experienced surgeons \citep{Jacobsen2019}. Moreover, the use of simulators has significantly reduced postoperative complication rates. Inexperienced surgeons had a 27.14$\%$ complication rate, compared to 12.86$\%$ for those with intermediate experience, with the difference being statistically significant \citep{Lucas2019}. Capsular rupture rates also decreased by 38$\%$ for surgical residents using the simulator \citep{Ferris2020}, and capsulorhexis errors were reduced by 68$\%$ after the simulator was introduced to training programs \citep{McCannel2013}.

\subsection{Prediction of Surgical Errors}

A significant portion of research on surgical error detection focuses on gesture error detection, often utilizing robotic data, particularly kinematic data \citep{9153443}, and occasionally relying on image data \citep{Sirajudeen2024}. Beyond robotics, most image-based studies prioritize real-time detection of adverse events in surgical videos, with bleeding being a primary target \citep{Eppler2023}. In the specific case of detecting surgical errors in cataract surgery, one study focuses on capsulorhexis-related events \citep{Morita2020}. The authors demonstrated that the system, trained to detect surgical errors in real-time, was capable of making predictions in advance. However, these predictions did not accurately localize the errors in the future, which made it impossible to generate relevant alerts.
Despite these advancements, there is a noticeable gap in research on predicting future surgical errors, which is essential for generating timely alerts and recommendations. To address this gap, this work aims to position itself within the field of event anticipation in videos, a domain known as action anticipation \citep{zhong2023surveydeeplearningtechniques}.

Action anticipation in deep learning typically involves analyzing temporal video segments to predict future actions or events. Video-based models can be used, as well as approaches that incorporate spatial and temporal encoders. While some studies forecast future events to support classification tasks using temporally conditioned models \citep{9577590}, others enhance predictions by incorporating additional data modalities, such as optical flow \citep{8099985} or domain-specific knowledge about relevant objects \citep{Zhong2022AnticipativeFF}. There are numerous applications, although few are found in the field of surgery, and they primarily focus on surgical phase anticipation \citep{boels2024suprasurgicalphaserecognition}.

\subsection{Domain Transfer and Adaptation}

An effective approach for domain adaptation is data transformation, such as histogram matching \citep{DIP}, which facilitates data homogenization. Furthermore, adversarial techniques, including Domain-Adversarial Neural Networks (DANN) \citep{ganin2015unsuperviseddomainadaptationbackpropagation} and their specialized adaptations for video data \citep{kim2021learningcrossmodalcontrastivefeatures}, as well as attention-based methods \citep{9305483}, demonstrate considerable potential. However, these adversarial methods often necessitate large batch sizes to perform optimally, as noted by \citep{yao2020largebatchsizetraining}.
Statistical discrepancy-based techniques, such as CORrelation ALignment (CORAL) \citep{sun2016deep} and Maximum Mean Discrepancy (MMD) \citep{gretton2008kernel}, focus on aligning feature distributions by minimizing statistical differences between domains. These methods are particularly effective for feature alignment and are adaptable for video data. We have selected these approaches as the foundation for the development of our proposed algorithm.

\subsubsection{Objectives of the study}

This work aims to demonstrate the feasibility of predicting surgical errors during capsulorhexis in real life.

We first developed two novel datasets to enable the prediction of surgical errors during capsulorhexis: the first consists of annotated capsulorhexis videos from a surgical simulator, capturing a range of error types, while the second is a dataset of real surgery videos, where a subset was annotated. The unannotated videos were used for unsupervised domain adaptation. These original datasets serve as the foundation for training and evaluating our models. 
Building on this, we implemented a real-time deep learning system that bridges the simulator’s final evaluation with real-time surgical error prediction. The simulator provides a variety of surgical error scenarios often absent in real videos, making it especially valuable for training a deep learning algorithm.
Additionally, we demonstrated that knowledge could be transferred from the simulator to real-world scenarios through unsupervised domain adaptation. Our approach leveraged the extensive data from the simulator and real surgical videos without requiring labor-intensive annotation to train an algorithm that generates real-time alerts. 

To our knowledge, this is the first attempt to transfer knowledge from simulators to real-world cataract surgeries.

\section{Dataset and annotation}

This section presents the acquisition of videos from the simulator and real surgeries, their preparation, preprocessing, and annotation.

\subsection{Data}

\subsubsection{Simulator data}
\label{ssec:ssec2_1_1}

The capsulorhexis module of the EyeSi\textsuperscript{\textregistered} Surgical cataract surgery simulator allows the operator to practice the capsulorhexis step of the cataract surgery.

A total of 422 exercises were collected. These were performed by 11 users of the simulator located at Brest University Hospital between May 2022 and May 2023.

The average duration for this exercise is 106.8 seconds, the standard deviation is 56.9 seconds, ranging from 25 to 222 seconds.

For each exercise, we recorded video using a single camera view (monocular), capturing frames every 33ms (at a 30Hz rate). Their spatial resolution is 800$\times$600 pixels.

\begin{figure}[!]
    \centering
    \includegraphics[width=0.48\textwidth]{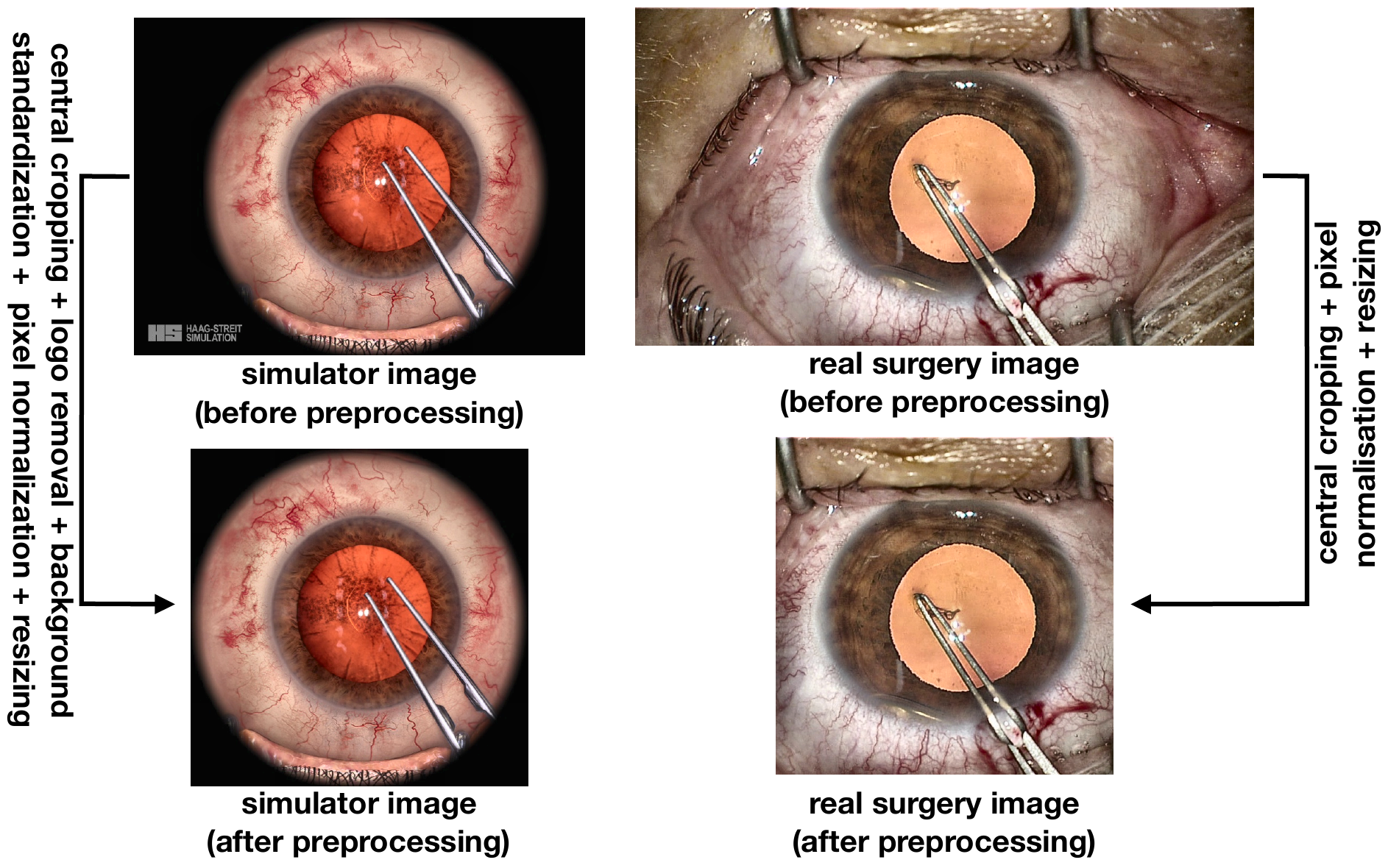}
    \caption{Steps and illustration of preprocessing results for images from simulator videos (left) and real surgery (right).}
    \label{fig:fig1}
\end{figure}

For video preprocessing, we cropped each image around the eye and ensured that all pixels representing the ocular border mask had a value of 0 through pixel normalization. The image size was then adjusted to 600$\times$600 pixels. The preprocessing result is shown in Fig.\ref{fig:fig1}.

Note that we also have access to the eye's three Euler angles across time, $\theta_{x}$, $\theta_{y}$, and $\theta_{z}$, which are used for annotating some surgical errors. These angles are expressed in a fixed reference frame and sampled at 30Hz.

\subsubsection{Real surgery data}

Between June and September 2023, we collected 107 monocular cataract surgery videos performed by five ophthalmologists at Brest University Hospital. Two surgeons had performed over 1000 surgeries, while the other three had completed fewer than 1000.
We have a single video for each patient.

We excluded 21 of these videos as they were either unusable (incomplete capsulorhexis, eye not in the camera frame) or depicted scenarios not represented in the simulator videos (e.g., white cataract).

The capsulorhexis step, our focus here, was manually extracted from those videos. In the following sections, when we refer to the available real surgical videos, we mean only the videos showing the capsulorhexis step.

\begin{figure}[!]
    \centering
    \includegraphics[width=0.48\textwidth]{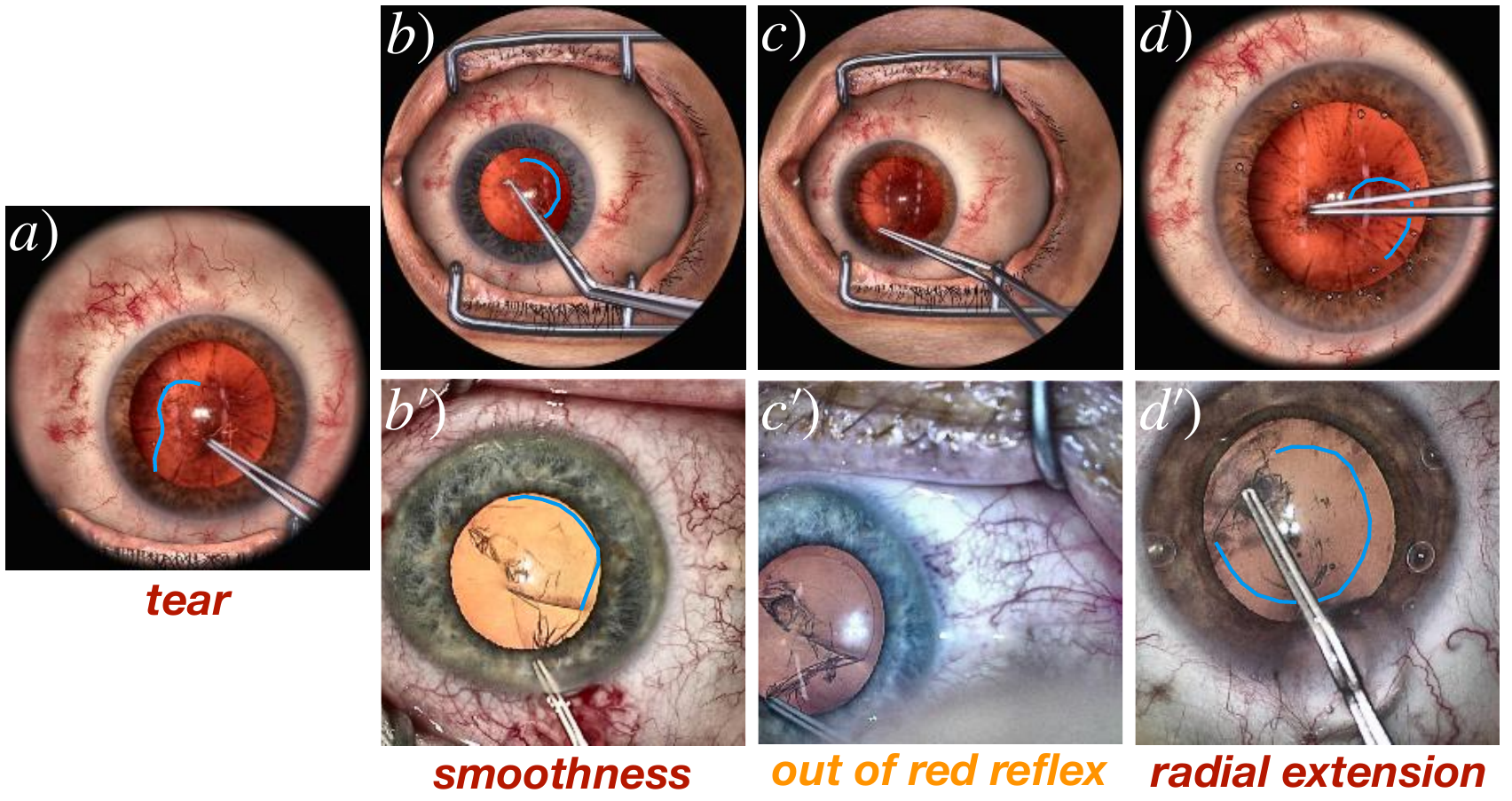}
    \caption{Illustrations of surgical errors on simulator video images and real surgical image: tear (a), smoothness (b, b'), out of red reflex (c, c'), and radial extension (d, d'). The capsulorhexis opening completed up to this point is highlighted in blue.}
    \label{fig:fig2}
\end{figure}

The average duration for performing capsulorhexis is 36.3 seconds, with a standard deviation of 20.4 seconds and ranging from 10 to 97 seconds. The videos are sampled at 30Hz, with a spatial resolution of 1280$\times$720. For preprocessing, the video images were normalized so that pixel values fell within the range of 0 to 1 and the image size was adjusted to 600$\times$600 pixels (Fig.\ref{fig:fig1}).

\subsection{Surgical errors}

\subsubsection{Description of simulator surgical errors at the video level}

The surgical simulator assesses performance and provides a score sheet at the end of the exercise, highlighting surgical errors in the video. The surgical errors evaluated by the simulator include:
\begin{itemize} 
\item \textit{operating without red reflex}: time spent with the eye out of red reflex. This metric is solely dependent on the eye and is not specific to the capsulorhexis step of cataract surgery.
\item \textit{smoothness}: roundness of the capsulorhexis, expressed as a percentage, 
\item \textit{radial extension}: maximum radius of the capsulorhexis in millimeters. 
\item \textit{tear}: the presence of a capsular tear, meaning a capsule rupture.
\end{itemize}

\begin{table}[]

\resizebox{0.48\textwidth}{!}{
\begin{tabular}{c|ll|ll|}
\cline{2-5}
\multicolumn{1}{l|}{}                   & \multicolumn{2}{c|}{simulator}                                                                                               & \multicolumn{2}{c|}{real surgery}                                                                                       \\ \cline{2-5} 
\multicolumn{1}{l|}{}                   & \multicolumn{1}{c|}{\begin{tabular}[c]{@{}c@{}}number of \\ surgical error\end{tabular}} & \multicolumn{1}{c|}{duration (s)} & \multicolumn{1}{c|}{\begin{tabular}[c]{@{}c@{}}number of\\ surgical error\end{tabular}} & \multicolumn{1}{c|}{duration (s)} \\ \hline
\multicolumn{1}{|c|}{\textit{smoothness}}        & \multicolumn{1}{c|}{760}                                                                 & \multicolumn{1}{c|}{$\times$}            & \multicolumn{1}{c|}{27}                                                                 & \multicolumn{1}{c|}{$\times$}        \\ \hline
\multicolumn{1}{|c|}{\textit{out of red reflex}} & \multicolumn{1}{c|}{756}                                                                 & \multicolumn{1}{c|}{712}                              & \multicolumn{1}{c|}{46}                                                                 & \multicolumn{1}{c|}{45}                            \\ \hline
\multicolumn{1}{|c|}{\textit{radial extension}}  & \multicolumn{1}{c|}{226}                                                                 & \multicolumn{1}{c|}{453}                               & \multicolumn{1}{c|}{9}                                                                  & \multicolumn{1}{c|}{6}                             \\ \hline
\multicolumn{1}{|c|}{\textit{tear}}              & \multicolumn{1}{c|}{44}                                                                  & \multicolumn{1}{c|}{$\times$}                                  & \multicolumn{1}{c|}{0}                                                                  & \multicolumn{1}{c|}{$\times$}                             \\ \hline

\end{tabular}}

\caption{\label{tab:tab1}Number of surgical errors and duration (in seconds) for the 422 annotated simulator videos and the 29 annotated real surgical videos.}

\end{table}

It is important to note that the occurrence of a surgical error does not necessarily indicate a failure; a capsulorhexis can still be deemed successful even if it is not perfect. Some surgical errors are less critical than others. For example, 100\% of capsulorhexis cases with a tear are considered failures, while this is not the case for other errors.

\subsubsection{Annotation of surgical errors over time}

An expert proceeded with the localization of surgical errors within the videos during a data annotation phase at the frame level to know exactly when surgical errors occur.

For annotation, we set aside videos that do not contain surgical errors according to the score sheets and annotated the remaining ones only.

Here are the error classes we defined, illustrated in Fig.\ref{fig:fig2} (a, b, c, d, and e):

\begin{itemize}  
\item The \textit{out of red reflex} (Fig.\ref{fig:fig2}$c$) error is automatically annotated using the eye's Euler angles at time $t$ if the empirically verified condition $\theta_{xt}^2 + \theta_{yt}^2 > \left(\cfrac{\pi}{12}\right)^2$ is met.
\item The initial frames at the start of a loss of circularity/regularity are annotated as containing a \textit{smoothness} error (Fig.\ref{fig:fig2}$b$)
\item When the capsulorhexis flap is too far from the center (Fig.\ref{fig:fig2}$d$): \textit{radial extension}. 
\item If a tear is present on an image (Fig.\ref{fig:fig2}$a$): \textit{tear} 
\end{itemize}

Regarding the statistics from the simulator, among the 422 videos, \textit{red reflex} errors occurred 756 times with a median duration of 0.67 seconds. \textit{Radial extension} errors occurred 226 times (1.73 seconds), \textit{tear} errors only 44 times, and \textit{smoothness} errors occurred 760 times.

\begin{figure}[!] 

\centering
\includegraphics[width=0.5\textwidth]{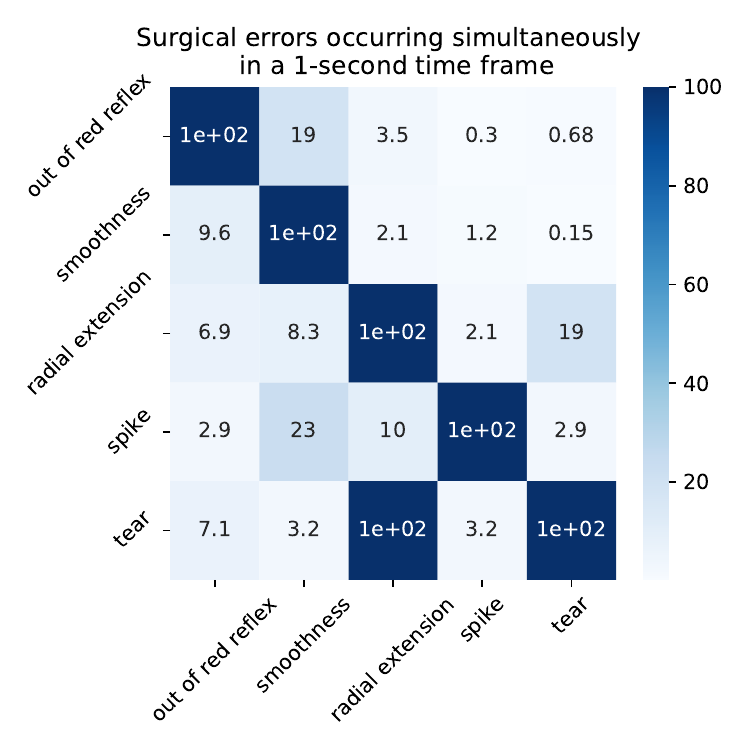}
\caption{Co-occurrence of surgical errors within a 1-second window on simulator data.}

\label{fig:fig3}
\end{figure}

Fig.\ref{fig:fig3} illustrates the surgical errors co-occurrences matrix $M$ between $\text{err}_A$ and $\text{err}_B$ in simulator data: 
\begin{align} 
M(\text{err}_A, \text{err}_B) = 100 \times \frac{\vert S_A \cap S_B \vert}{\vert S_A \vert}.
\label{eq:eq10}
\end{align}

\noindent where $S_A$ and $S_B$ represent sequences of binary labels of length $T$ (total duration of simulator videos in second), with $S_A, S_B \in \{0, 1\}^T \times \{0, 1\}^T$ indicate the presence ($1$) or absence ($0$) of an error at each 1-second time sequence. $M(\text{err}_A, \text{err}_B)$ denote the proportion of the error sequence $S_A$ that coincides with the errors in $S_B$.

The co-occurrence of errors is generally low, as the occurrence of one error does not necessarily lead to another, except for \textit{tear} $\rightarrow$ \textit{radial extension}. The asymmetry of the matrix shows that the reverse is not true.

While we do not have score sheets for the real surgical videos, annotations are performed on a subset of 29 videos. Surgical errors in real surgeries are much less frequent — approximately 23 times fewer. 

The least observed surgical error in the simulator videos (\textit{tear}) is absent in annotated real surgeries, as shown in Tab.\ref{tab:tab1}. The median duration for the \textit{red reflex} error is 0.73 seconds, while it is 0.70 seconds for the \textit{radial extension}. Some annotations are illustrated in Fig.\ref{fig:fig2} (b', c', and d').

\subsection{Data split}

Simulator videos are split into 60/20/20\% for the training/validation/test sets. We ensured that the surgical error content was evenly distributed across each set.

The split for the dataset of real surgical videos is 60/28/12\%. This was done on unannotated videos, ensuring each split contained the same proportion of videos recorded on the same day. The split was necessary to account for slight visual and skill variations in the videos due to different camera settings and the surgeon performing the procedure. This ensured that the diversity of video appearances and surgeon expertise was evenly distributed across the datasets.

The 29 validation and test videos of the real video dataset were annotated afterward.

The training, validation, and test sets for simulator and surgical videos are decomposed into video snippets for error prediction.

\section{Methods}

\subsection{Decomposition into video snippets}

Simulator and surgical videos are segmented into snippets for the training, validation, and test sets. To simplify, we did not include the names of the videos from which each snippet is derived in the notation. Instead, each snippet is indexed by $k$.

Let $A^{(k)}_{[t - D_k : t]}$ denote the prior video snippet at time $t$, lasting $D_k$ seconds, and composed of $L_k$ frames. For labeled videos, we also define $P^{(k)}_{]t + \Delta T_k : t + \Delta T_k + 1]}$ as the posterior sequence, starting $\Delta T_k$ seconds (the prediction horizon) after the end of $A^{(k)}_{[t - D_k : t]}$ and lasting for 1 second. We denote $S_k \in \mathbb{N}^*$ such that $S_k - 1$ represents the number of frames skipped between two consecutive frames in the original video. This parameter determines the temporal downsampling applied to create the video snippet. Since the base sampling on the simulator is 30 frames per second, the temporal duration $D_k$ of $A^{(k)}_{[t - D_k: t]}$ in seconds is:

\begin{align}
D_k = \cfrac{L_k \times S_k}{30}
\label{eq:eq2}
\end{align}

Thus, by choosing $L_k$ and $S_k$, we control the extent of the history to predict surgical errors. For example, by choosing $L_k = 30$ and $S_k = 1$ or $L_k = 10$ and $S_k = 3$, we have $D_k = 1$ s. The prior video snippet is observed for prediction by the algorithm, while the posterior sequence is used to define the label (surgical error class) $\bm{l}_{\Delta T_k}$, where $\bm{l}_{\Delta T_k} \in \{0, 1\}^{C + 1}$. In our study, $C = 5$ represents the number of surgical classes, with the addition of +1 corresponding to one additional class for all surgical errors combined. Specifically, we aim to predict from the video snippet $A^{(k)}_{[t - D_k : t]}$ whether a surgical error will occur across the sequence $P^{(k)}_{]t + \Delta T_k : t + \Delta T_k + 1]}$.

Concretely:
\begin{align}
\bm{l}_{\Delta T_k}[j] = \left\{
    \begin{array}{ll}
        1 & \text{if a surgical error of class } j \text{ occurs} \\
          & \text{across the sequence } P^{(k)}_{]t + \Delta T_k, t + \Delta T_k + 1]} \\
        0 & \text{otherwise}
    \end{array}
\right.
\label{eq:eq3}
\end{align}

\begin{figure}[!]
\centering
\includegraphics[width=0.485\textwidth]{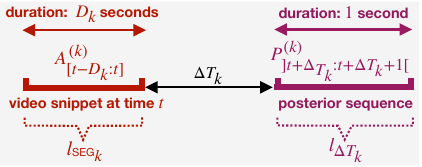}
\caption{Illustration of the concepts composing our dataset: $A^{(k)}_{[t - D_k : t]}$ the video snippet observed by the model of duration $D_k$ seconds, which defines the current label $\bm{l}_{\seg_k}$, and the posterior sequence $P^{(k)}_{]t + \Delta T_k : t + \Delta T_k + 1]}$, of duration 1 second used to define the prediction label $\bm{l}_{\Delta T_k}$. Note that the data is indexed by $k$ and that the original video notation is not involved.}
\label{fig:fig4}
\end{figure}

We also classify the current error in the observed video snippet by defining the current label $\bm{l}_{\seg_k} \in \{0, 1\}^{L_k, C + 1}$. This label corresponds to the surgical error content in the observed video snippet, ensuring that the extracted features during training focus more on the current error information rather than other irrelevant details (Fig.\ref{fig:fig4}). We set $\Delta T_k = T$ with $T \in [1, 5]$ seconds. We arbitrarily shifted the sequences by $1/15$ second for simulator data to create a dataset with diverse sequences.
Additionally, we used markers for the training set and validation set to select every third clip without surgical errors, which helped balance the dataset. For the surgical videos, which are fewer in number, we consistently shifted by $1/30$. Tab.\ref{tab:tab2} illustrates the number of video snippets associated with each type of surgical error for a training dataset constructed with $L_k = 10$ and $S_k = 3$. Note that we ensured to have the same number of video snippets and the same proportion of surgical errors for the considered configurations $(L_k, S_k)$ to ensure fair comparisons between various settings.

\begin{table}[]

\centering
\begin{tabular}{c|l|l|}
\cline{2-3}
\multicolumn{1}{l|}{}                   & \multicolumn{1}{c|}{\begin{tabular}[c]{@{}c@{}}number of snippet\end{tabular}} & \multicolumn{1}{c|}{proportion ($\%$)} \\ \hline
\multicolumn{1}{|c|}{\textit{smoothness}}        & \multicolumn{1}{|c|}{18,787}                                                                                               & \multicolumn{1}{|c|}{21}                                   \\ \hline
\multicolumn{1}{|c|}{\textit{out of red reflex}} & \multicolumn{1}{|c|}{10,361}                                                                                               & \multicolumn{1}{|c|}{12}                                 \\ \hline
\multicolumn{1}{|c|}{\textit{radial extension}}  & \multicolumn{1}{|c|}{5,443}                                                                                               & \multicolumn{1}{|c|}{6.2}                                  \\ \hline
\multicolumn{1}{|c|}{\textit{tear}}              & \multicolumn{1}{|c|}{294}                                                                                                & \multicolumn{1}{|c|}{0.34}                                  \\ \hline
\multicolumn{1}{|c|}{\textit{no error}}          & \multicolumn{1}{|c|}{50,893}                                                                                              & \multicolumn{1}{|c|}{58.8}                                 \\ \hline
\multicolumn{1}{|c|}{total}             & \multicolumn{1}{|c|}{86,468}                                                                                              & \multicolumn{1}{c|}{100}               \\ \hline
\end{tabular}
\caption{\label{tab:tab2}Distribution of surgical errors at the video snippet scale for the training dataset generated with $L_k=10$, $S_k=3$ and $\Delta T_k = 1$ (simulator data).}
\end{table}

\subsection{Surgical error prediction from simulator data}

In this subsection, we present the prediction model that will be trained in a supervised manner using data from the simulator.

\subsubsection{Algorithm for surgical error prediction}

We aim to implement a real-time analysis system. Therefore, we develop a conditional algorithm to make predictions with a customizable prediction horizon.

We denote by $\mathcal{F}$ the spatial encoder model, $\mathcal{M}$ the temporal encoder model, $\mathcal{C}^{(c)}$ two classification layers for the current $(c)$ classification, $\mathcal{C}^{(p)}$ two classification layers for prediction $(p)$, $\texttt{att}$ an attention layer and $\TE$ two fully-connected layers. 

Thus, we have:
\begin{itemize}
\item $\bm{SR}_{k} = \mathcal{F}(A^{(k)}_{[t - D_k: t]})$. It corresponds to applying $\mathcal{F}$ to each element of the sequence $A^{(k)}_{[t - D_k: t]}$.

\item $\bm{TR}_{k} = \mathcal{M}(\bm{SR}_{k})$
\item $\bm{CC}_{k} = \mathcal{C}^{(c)}(\bm{TR}_{k})$ and $\bm{CC}_{k}^{(\texttt{att})} =  \hat{\bm{l}}_{\seg_k} = \texttt{att}(\bm{CC}_{k})$
\item $\bm{TR}_{k}^{(+)} = \concat(\bm{TR}_{k}, \bm{CC}_{k}^{(\texttt{att})}, \TE(\Delta T_k))$
\item $\hat{\bm{l}}_{\Delta T_k} = \bm{FP}_{k} = \mathcal{C}^{(p)}(\bm{TR}_{k}^{(+)})$
\end{itemize}
\noindent with $\bm{SR}_{k} \in \mathbb{R}^{1, L_k, I}$ being the spatial representation with $I$ as the latent representation size of the images.

$\bm{TR}_{k} \in \mathbb{R}^{1, S}$ is the temporal representation of the data with $S$ being the chosen size of this latent representation. $\bm{CC}_{k} \in \mathbb{R}^{1, L_k, C + 1}$ (current classification) is the surgical error classification for each image of the observed video snippet. $\bm{CC}_{k}^{(\texttt{att})} \in \mathbb{R}^{1, C + 1}$ then corresponds to the average surgical error (averaged by an attention mechanism) of the previous sequence. Thus, $\bm{TR}_k$ and $\bm{CC}_k$ can be seen as information relative to the previous sequence.

$\TE(\Delta T_k) \in \mathbb{R}^{1, T}$ corresponds to a learned temporal representation of the prediction horizon $\Delta T_k$.

$\bm{TR}_{k}^{(+)} \in \mathbb{R}^{1, S + C + 1 + T}$ is the concatenation of the temporal representation, the predicted surgical error content of the observed sequence, and the prediction horizon to customize the future prediction $\bm{FP}_{k} \in \mathbb{R}^{1, C + 1}$. We trained a conditional model to avoid training multiple models for predictions across different prediction horizons $\Delta T_k$.

\begin{figure*}[!]
\centering
\centering
\includegraphics[width=1\textwidth]{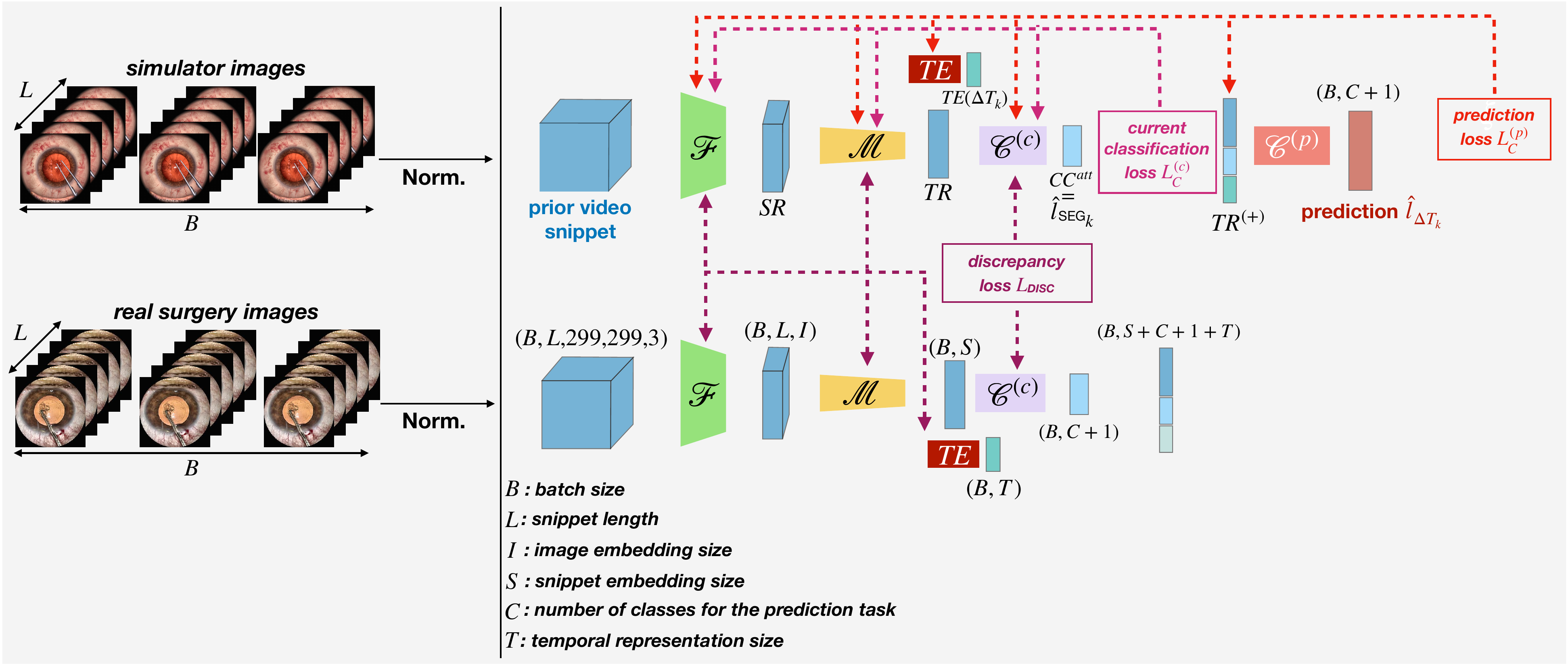}
\caption{\label{fig:fig5}Simplified diagram of the unsupervised domain adaptation approach based on distribution alignment. The dotted arrows represent the backpropagation of the gradient.}

\end{figure*}

The upper part of Fig.\ref{fig:fig5} shows the different parts of the deep learning algorithm used for error prediction in surgical procedures based on the simulator.

Note that for the sake of clarity, we have fixed the batch size to $B=1$ here.

\subsubsection{Loss Function}
\label{sub:lossF}

To train the model, we minimize the empirical risk defined by a cross-entropy loss function with two terms:

\begin{multline}
\mathcal{L}_{C}(.) = \lambda_1 \mathcal{L}_{C}^{(c)}(W_{\cal{F}}, W_{\cal{M}}, W_{\att}, W_{\mathcal{C}^{(c)}}) \\ + \mathcal{L}_{C}^{(p)}(W_{\cal{F}}, W_{\cal{M}}, W_{\att}, W_{\mathcal{C}^{(c)}}, W_{\TE}, W_{\mathcal{C}^{(p)}}) \\
= - \frac{\lambda_1}{B} \frac{1}{B} \sum_{k=1}^B \sum_{c=1}^C \bm{l}_{\seg_k}[c] \log(\hat{\bm{l}}_{\seg_k}[c])
- \sum_{k=1}^B \sum_{c=1}^C \bm{l}_{\Delta T_k}[c] \log(\hat{\bm{l}}_{\Delta T_k}[c])
\end{multline}

The first term $\mathcal{L}_{C}^{(c)}(W_{\cal{F}}, W_{\cal{M}}, W_{\att}, W_{\mathcal{C}^{(c)}})$ is the current classification loss,  where $W_\mathcal{F}$ corresponds to the learned weights of the model $\mathcal{F}$. It is associated with the classification of the surgical error that occurred in the previous sequence. This term enables the model to learn the observed sequence's surgical error content and generate a richer representation to guide the final classification. The second term corresponds to the prediction made within the prediction horizon $\Delta T_k$.

\subsection{Unsupervised domain adaptation}

After addressing the prediction of surgical errors using simulator videos, the focus now shifts to adapting to real surgical videos without available annotations for training. This challenge necessitates unsupervised domain adaptation techniques to predict surgical errors in real videos by leveraging knowledge from annotated simulator data while adapting the differences between the two domains.

\subsubsection{Data homogenization}

Homogenizing the data is the most straightforward strategy when considering knowledge transfer from one domain to another. The goal is to directly bring the target domain data closer to the source domain.

\begin{figure}[!]
    \centering
    \includegraphics[width=0.485\textwidth]{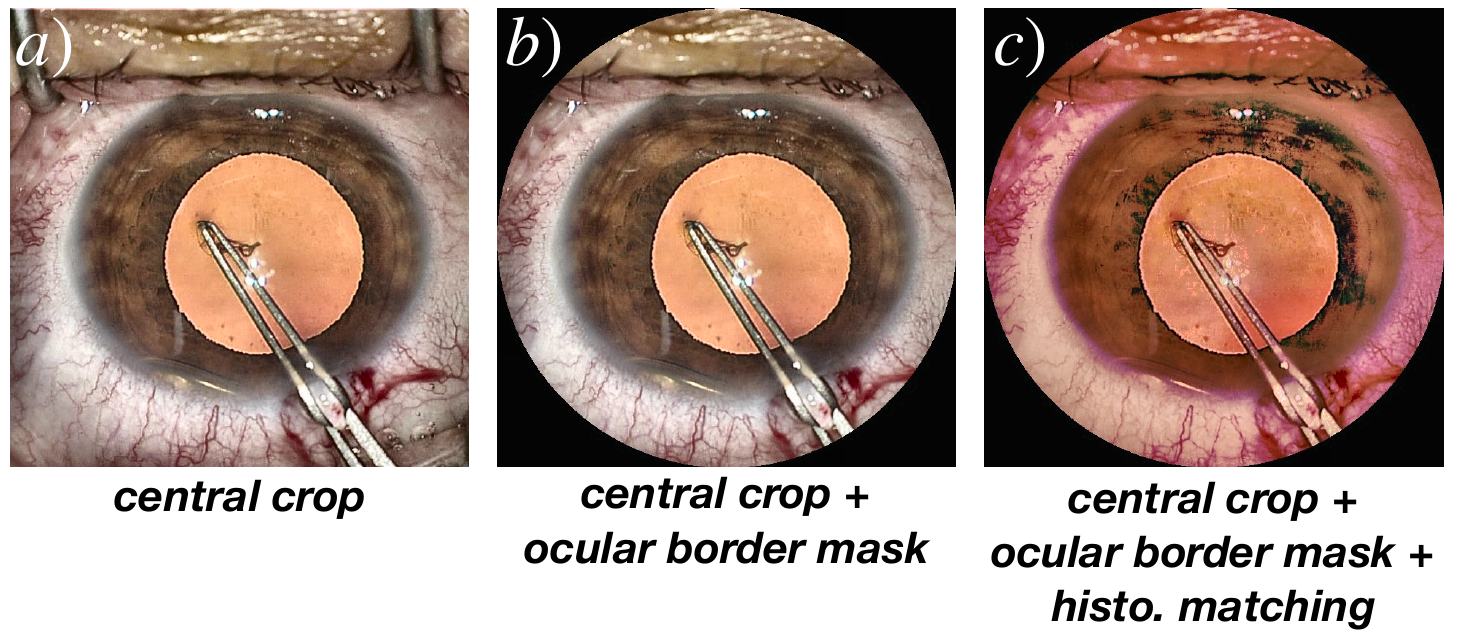}
    \caption{Illustrations of images from real surgical videos: central crop from a simulator image (a), images with ocular border mask (b), and images with ocular border mask and histogram matching (c).}
    \label{fig:fig6}
\end{figure}

We applied several preprocessing steps to the images from real surgeries, as illustrated in Fig.\ref{fig:fig6}.

\begin{itemize} \item \textit{central crop}: Extracting a central square crop from the image. This is the standard preprocessing technique mentioned in Sect.\ref{ssec:ssec2_1_1}. \item \textit{ocular border mask}: Adding black pixels around the central square crop to align the histograms of the images (simulator images contain many black pixels). \item \textit{ocular border mask + histogram matching}: Histogram matching is an image processing technique used to adjust the pixel intensity distribution of one image to match the histogram of another \citep{DIP}. This method is beneficial to standardize brightness, contrast, or color distribution between images. The intuition behind applying this technique is to align the color tones of real surgical video frames with those of the simulator, reducing variance and simplifying the task for the network. Several reference images (10) were selected based on iris color variability and magnification level to ensure a straightforward and effective computation. \end{itemize}

This strategy does not include any additional training. A model trained on simulator data will be directly evaluated on the transformed video snippets from the real surgery validation set.

\subsubsection{Feature alignment}
\label{sub:FeaAl}

As the variance of the images at the pixel level is very large, we have considered a transfer approach that passes through the latent space.

This is an unsupervised approach that does not exploit the labels of the classification on the simulator.

The idea is to align the representations of the video snippets from the simulator using a discrepancy loss. We chose to perform this alignment with a statistical method that brings the distributions closer by considering a batch of video snippets.

In our study, we have considered the CORrelation ALignment (CORAL) method \citep{sun2016deep}, which aims to align the covariance matrices of the extracted features for the source and target data by minimizing their difference.
\begin{align}
\mathcal{L}_{\coral}(.) = \cfrac{1}{4S^2} || C_S - C_T ||^2_F
\end{align}
\noindent where $||...||^2_F$ represents the Frobenius norm and with:
\begin{align}
C_S = \cfrac{1}{B - 1} \bigg(\bm{TR}_{(S)}^T \bm{TR}_{(S)} - \cfrac{1}{B} (\bm{1}^T \bm{TR}_{(S)})^T (\bm{1}^T \bm{TR}_{(S)})\bigg) 
\end{align}
\begin{align}
C_T = \cfrac{1}{B - 1} 
\bigg(\bm{TR}_{(T)}^T \bm{TR}_{(T)} - \cfrac{1}{B} (\bm{1}^T \bm{TR}_{(T)})^T (\bm{1}^T \bm{TR}_{(T)})\bigg)
\end{align}
\noindent the covariance matrices estimated from the batch of $B$ features of the video snippets from the source (simulator) $_{(S)}$ and target (real surgery) $_{(T)}$ domains, respectively.

This loss can be seen as an alignment constraint. In parallel, the classification task is performed solely on the simulator data.

We also considered in our study the Maximum Mean Discrepancy (MMD), which is a measure between latent representations involving a kernel \citep{gretton2008kernel}.

\begin{align*}
\mathcal{L}_{\mmd}^{2}(.) & = \frac{1}{B(B-1)} \sum_{i} \sum_{j \neq i} \varphi(\bm{TR}_{(S)}[i], \bm{TR}_{(S)}[j]) \\
        &- 2 \frac{1}{B^2} \sum_{i} \sum_{j} \varphi(\bm{TR}_{(S)}[i], \bm{TR}_{(T)}[j]) \\
        &+ \frac{1}{B(B-1)} \sum_{i} \sum_{j \neq i} \varphi(\bm{TR}_{(T)}[i], \bm{TR}_{(T)}[j])
\end{align*}

with $\varphi(\mathbf{x}, \mathbf{y}) = \exp \left(\frac{- \Vert \mathbf{x} - \mathbf{y} \Vert^{2}}{2\sigma^{2}}\right)$.

We also compared these losses to the Mean Squared Error (MSE):

\begin{align*}
\mathcal{L}_{\mse}(.) & = \frac{1}{B} \sum_{i} (\bm{TR}_{(S)}[i] - \bm{TR}_{(T)}[i])^2
\end{align*}

The overall loss is therefore:

\begin{center}
    $\mathcal{L}(.) = \mathcal{L}_C(W_{\cal{F}}, W_{\cal{M}}, W_{\att}, W_{\mathcal{C}^{(c)}}, W_{\TE}) + \lambda_2 \mathcal{L}_{\disc}(W_{\cal{F}}, W_{\cal{M}}, W_{\att}, W_{\mathcal{C}^{(c)}}, W_{\TE}, W_{\mathcal{C}^{p)}})$
\end{center}

where we set $\mathcal{L}_{\disc}$ as either $\mathcal{L}_{\coral}$, $\mathcal{L}_{\mmd}$, or $\mathcal{L}_{\mse}$.

Fig.\ref{fig:fig5} illustrates the training procedure for feature alignment. 

\subsubsection{Pre-training}

Self-supervised learning is particularly useful in domain adaptation when dealing with multiple domains, as it enables the definition of tasks that extract feature sets common to both domains. This approach can facilitate subsequent classification tasks and potentially enhance performance \citep{misra2016shufflelearnunsupervisedlearning}.

We defined two self-supervised learning tasks, also known as pretext tasks, to pre-train the model. 

These tasks are performed simultaneously on both the simulator data and real surgery data, allowing the model to learn shared representations from both domains and improve its ability to adapt to domain-specific challenges.

\begin{itemize}
    \item \textbf{Frame Order Task}: This task is formulated as a binary classification problem where the model predicts whether the frames in the input video snippet are in the correct temporal order. The associated cross-entropy loss is $\mathcal{L}_{C}^{(p)}(W_{\cal{F}}, W_{\cal{M}}, W_{\att}, W_{\mathcal{C}^{(p)}})$ with $\mathcal{C}^{(p)}$ represents the classification layer for this pretext task.

    \item \textbf{Image Reconstruction Task}: This task is framed as an image reconstruction problem. The model is trained to reconstruct a randomly cropped quarter of an image after a prediction horizon $\Delta T_k$. The objective is to minimize the mean squared error (MSE) between the predicted and target images. The loss function is defined as $\mathcal{L}_{\mse}(W_{\cal{F}}, W_{\cal{M}}, W_{\att}, W_{\mathcal{C}^{(c)}}, W_{\TE}, W_{\mathcal{D}} )$ guiding the model to accurately predict the image across the temporal gap. The decoder, $\cal{D}$, is an LSTM-CNN symmetric to the encoder.
\end{itemize}

\section{Experiments and results}

\subsection{Evaluation metric}

The Area Under the ROC Curve (AUC) is chosen as the evaluation metric for this study.

\subsection{Training settings}

We implemented the algorithms using the PyTorch framework and utilized the Adam optimizer with a learning rate of $5 \times 10^{-5}$ for the spatial encoder $5 \times 10^{-4}$ for the rest.

For training the model to predict surgical errors using only the simulator data, we employed batches of size $B = 32$ snippets to minimize empirical risk with gradient accumulation. We set $\lambda_1 = 0.1$ (see \ref{sub:lossF}) and trained for 10 epochs, employing early stopping based on overall AUC to prevent overfitting.

For domain adaptation, and computational efficiency, we used a batch size of $B = 8$ and trained for 10 epochs with $\lambda_2 = 1$ (see \ref{sub:FeaAl}).

Given the significant variability in the results from validation sets, we conducted five separate training runs. We selected the best outcomes based on their performance on the validation set.

\subsection{Surgical error prediction on simulator data}

For this study, we simplified the setup by setting the previous sequence to a duration of $D_k = 1$ second, with $L_k = 10$ frames, corresponding to a sampling rate of $S_k = 3$ samples/s.

Unless otherwise specified, the prediction horizon was set to 1 second for training and inference.

We evaluated multiple spatial and temporal encoders.
Tab.\ref{tab:tab3} presents the comparative study results. The inference time shown is the average time in milliseconds to compute $\bm{SR}_k$ over 1000 inferences. An asterisk (*) denotes spatial encoders pre-trained on images of different sizes than those used in our study (fourth column).

For temporal encoders, we used the 3 best configurations without pre-training:

\begin{itemize}
    \item LSTM: 2 layers, 256 hidden units,
    \item 1D CNN: dilated causal convolution \citep{oord2016wavenetgenerativemodelraw}, 
    \item Transformer encoder: 3 layers, 8 attention heads, and a 256-dimensional embedding.
\end{itemize}

\begin{table*}[]

\centering
\resizebox{\textwidth}{!}{

\begin{tabular}{rllllllll}
\cline{7-9}
\multicolumn{1}{c}{}                                     & \multicolumn{1}{c}{}                         & \multicolumn{1}{c}{}                                                                       & \multicolumn{1}{c}{}                                                       & \multicolumn{1}{c}{}                                                     & \multicolumn{1}{c|}{}                                                                       & \multicolumn{3}{c|}{Temporal encoder}                                                            \\ \hline
\multicolumn{1}{|c|}{Spatial encoder}                    & \multicolumn{1}{c|}{Version}                 & \multicolumn{1}{c|}{\begin{tabular}[c]{@{}c@{}}Number of \\ parameters\\ (M)\end{tabular}} & \multicolumn{1}{c|}{\begin{tabular}[c]{@{}c@{}}Image \\ size\end{tabular}} & \multicolumn{1}{c|}{\begin{tabular}[c]{@{}c@{}}FLOPs\\ (G)\end{tabular}} & \multicolumn{1}{c|}{\begin{tabular}[c]{@{}c@{}}Average inference \\ time (ms)\end{tabular}} & \multicolumn{1}{c|}{LSTM}     & \multicolumn{1}{c|}{1D-CNN}    & \multicolumn{1}{c|}{Transformer} \\ \hline
\multicolumn{1}{|r|}{\multirow{2}{*}{ResNet}}            & \multicolumn{1}{l|}{50}                      & \multicolumn{1}{l|}{\multirow{2}{*}{25.5}}                                                 & \multicolumn{1}{l|}{224}                                                   & \multicolumn{1}{l|}{4}                                                   & \multicolumn{1}{l|}{7.2 $\pm$ 0.04}                                                             & \multicolumn{1}{l|}{0.749 $\pm$ 0.006 } & \multicolumn{1}{l|}{0.746 $\pm$ 0.005} & \multicolumn{1}{l|}{0.749 $\pm$ 0.006}    \\ \cline{2-2} \cline{4-9} 
\multicolumn{1}{|r|}{}                                   & \multicolumn{1}{l|}{50*}                     & \multicolumn{1}{l|}{}                                                                      & \multicolumn{1}{l|}{600}                                                   & \multicolumn{1}{l|}{29}                                                  & \multicolumn{1}{l|}{7.5 $\pm$ 0.06}                                                             & \multicolumn{1}{l|}{0.789 $\pm$ 0.005 } & \multicolumn{1}{l|}{0.768 $\pm$ 0.007} & \multicolumn{1}{l|}{0.759 $\pm$ 0.007}    \\ \hline
\multicolumn{1}{|r|}{\multirow{2}{*}{Inception\_ResNet}} & \multicolumn{1}{l|}{v2}                      & \multicolumn{1}{l|}{\multirow{2}{*}{56.4}}                                                 & \multicolumn{1}{l|}{299}                                                   & \multicolumn{1}{l|}{13}                                                  & \multicolumn{1}{l|}{30.1 $\pm$ 2.2}                                                             & \multicolumn{1}{l|}{0.784 $\pm$ 0.005 } & \multicolumn{1}{l|}{0.754 $\pm$ 0.007} & \multicolumn{1}{l|}{0.755 $\pm$ 0.006}    \\ \cline{2-2} \cline{4-9} 
\multicolumn{1}{|r|}{}                                   & \multicolumn{1}{l|}{v2*}                     & \multicolumn{1}{l|}{}                                                                      & \multicolumn{1}{l|}{600}                                                   & \multicolumn{1}{l|}{56}                                                  & \multicolumn{1}{l|}{42.4 $\pm$ 0.03}                                                            & \multicolumn{1}{l|}{0.796 $\pm$ 0.005} & \multicolumn{1}{l|}{\textbf{0.783 $\pm$ 0.007}} & \multicolumn{1}{l|}{\textbf{0.802 $\pm$ 0.005}}    \\ \hline
\multicolumn{1}{|r|}{\multirow{3}{*}{EfficientNet}}      & \multicolumn{1}{l|}{b0ns}                    & \multicolumn{1}{l|}{\multirow{2}{*}{5}}                                                    & \multicolumn{1}{l|}{224}                                                   & \multicolumn{1}{l|}{0.4}                                                 & \multicolumn{1}{l|}{8.1 $\pm$ 1.9}                                                              & \multicolumn{1}{l|}{0.776 $\pm$ 0.006 } & \multicolumn{1}{l|}{0.743 $\pm$ 0.005}   & \multicolumn{1}{l|}{0.732 $\pm$ 0.006}    \\ \cline{2-2} \cline{4-9} 
\multicolumn{1}{|r|}{}                                   & \multicolumn{1}{l|}{b0ns*}                     & \multicolumn{1}{l|}{}                                                                      & \multicolumn{1}{l|}{600}                                                   & \multicolumn{1}{l|}{3}                                                   & \multicolumn{1}{l|}{9.6 $\pm$ 2.5}                                                              & \multicolumn{1}{l|}{\textbf{0.812 $\pm$ 0.005}} & \multicolumn{1}{l|}{0.768 $\pm$ 0.006} & \multicolumn{1}{l|}{0.780 $\pm$ 0.005}    \\ \cline{2-9} 
\multicolumn{1}{|r|}{}                                   & \multicolumn{1}{l|}{b7ns}                    & \multicolumn{1}{l|}{66}                                                                    & \multicolumn{1}{l|}{600}                                                   & \multicolumn{1}{l|}{38}                                                  & \multicolumn{1}{l|}{54.6 $\pm$ 0.05}                                                            & \multicolumn{1}{l|}{0.784 $\pm$ 0.006 } & \multicolumn{1}{l|}{0.774 $\pm$ 0.005} & \multicolumn{1}{l|}{0.795 $\pm$ 0.005}    \\ \hline
\multicolumn{1}{|r|}{\multirow{2}{*}{InceptionNet}}    & \multicolumn{1}{l|}{v3}                      & \multicolumn{1}{l|}{\multirow{2}{*}{24}}                                                   & \multicolumn{1}{l|}{299}                                                   & \multicolumn{1}{l|}{6}                                                   & \multicolumn{1}{l|}{12.2 $\pm$1.2}                                                             & \multicolumn{1}{l|}{0.766 $\pm$ 0.006} & \multicolumn{1}{l|}{0.751 $\pm$ 0.005} & \multicolumn{1}{l|}{0.773 $\pm$ 0.006}    \\ \cline{2-2} \cline{4-9} 
\multicolumn{1}{|r|}{}                                   & \multicolumn{1}{l|}{v3*}                     & \multicolumn{1}{l|}{}                                                                      & \multicolumn{1}{l|}{600}                                                   & \multicolumn{1}{l|}{24}                                                  & \multicolumn{1}{l|}{13.6 $\pm$ 1.7}                                                             & \multicolumn{1}{l|}{0.789 $\pm$ 0.005 } & \multicolumn{1}{l|}{0.751 $\pm$ 0.007} & \multicolumn{1}{l|}{0.753 $\pm$ 0.006}    \\ \hline
\multicolumn{1}{|r|}{\multirow{2}{*}{ViT}}               & \multicolumn{1}{l|}{vit\_base\_patch16\_224} & \multicolumn{1}{l|}{87}                                                                    & \multicolumn{1}{l|}{224}                                                   & \multicolumn{1}{l|}{18}                                                  & \multicolumn{1}{l|}{10.8 $\pm$ 1.8}                                                             & \multicolumn{1}{l|}{0.752 $\pm$ 0.006} & \multicolumn{1}{l|}{0.731 $\pm$ 0.007} & \multicolumn{1}{l|}{0.734 $\pm$ 0.006}    \\ \cline{2-9} 
\multicolumn{1}{|r|}{}                                   & \multicolumn{1}{l|}{vit\_base\_patch16\_384} & \multicolumn{1}{l|}{87}                                                                    & \multicolumn{1}{l|}{384}                                                   & \multicolumn{1}{l|}{56}                                                  & \multicolumn{1}{l|}{12.8 $\pm$ 1.7}                                                             & \multicolumn{1}{l|}{0.779 $\pm$ 0.006} & \multicolumn{1}{l|}{0.758 $\pm$ 0.005} & \multicolumn{1}{l|}{0.764 $\pm$ 0.006}    \\ \hline

                                                         \\ \cline{7-9} 
\multicolumn{6}{r|}{}                                                                                                                                                                                            & \multicolumn{3}{c|}{None}                                                                        \\ \hline
\multicolumn{1}{|r|}{X3D}                                & \multicolumn{1}{l|}{M}                     & \multicolumn{1}{l|}{4}                                                                     & \multicolumn{1}{l|}{224}                                                   & \multicolumn{1}{l|}{5}                                                   & \multicolumn{1}{l|}{19 $\pm$ 1.3}                                                                     & \multicolumn{3}{l|}{0.762 $\pm$ 0.008
}                                                                    \\ \hline
\multicolumn{1}{l}{}                                     &                                                                 
\end{tabular}}

\caption{\label{tab:tab3}Overall results (AUC) on the validation simulator dataset for various temporal and spatial encoders, with $L_k = 10$, $S_k = 3$, and $\Delta T_k = 1$ s. Results are presented as the Mean (95$\%$ CI), derived from the DeLong test. Significant results are highlighted in bold.}

\end{table*}

In our comparative study, 2D CNN models surpassed vision Transformers (with spatial encoder) and 3D CNN model \citep{feichtenhofer2020x3dexpandingarchitecturesefficient}. The best-performing spatial encoders were:

\begin{itemize}
    \item Inception$\_$ResNetv2 pre-trained on ImageNet using 299$\times$299 pixel images, further trained for surgical error prediction using 600$\times$600 pixel images 
    \item EfficientNetB0 pre-trained on noisy student using 224$\times$224 pixel images, further trained for surgical error prediction using 600$\times$600 pixel images
\end{itemize}

However, EfficientNetB0 achieved the highest AUC of 0.812 with 600$\times$600 pixel images on the validation data, with statistical significance (p-value = 1.89e-05, DeLong test \citep{DeLong1988}).

In general, the LSTM outperformed other temporal encoders or achieved similar results, except for the Transformer when using EfficientNetB7 (AUC = 0.795 compared to 0.784, with a p-value of 1.55e-05), and InceptionV3 with an image size of 299$\times$299 (AUC = 0.773 compared to 0.766, p-value of 1.77e-05).

In all cases, the results are better when error prediction is performed using 600$\times$600 images compared to smaller sizes (224$\times$224, 299$\times$299, 384$\times$384 pixels), even when the model was pre-trained with smaller image sizes. This underscores the importance of larger image sizes for capturing comprehensive eye movement and information related to the capsulorhexis flap.

Note that the best result with a size smaller than 600$\times$600 pixels was achieved using Inception$\_$ResNetv2 and LSTM, which was pre-trained on ImageNet with a size of 299$\times$299 pixels and trained for surgical error prediction with 299$\times$299 pixels, resulting in an AUC of 0.784. Tab.\ref{tab:tab4} presents the results from the second comparative study. We maintained the best configuration (EfficientNetB0 + LSTM, 600$\times$600 pixel images) and varied sequence hyperparameters ($L_k$ and $S_k$). The prediction horizon remained fixed at 1 second for inference. Note that, here, the inference time corresponds to the average time to compute future predictions for a batch size of 1.

\begin{table}[]

\centering
\begin{tabular}{|l|l|l|l|l|}
\hline
$D_k$ (s)                   & $L_k$ & $S_k$ & inference time (ms) & AUC (overall) \\ \hline
1/30                   & 1    & 1    & 6.8 $\pm$ 0.6           & 0.734 $\pm$ 0.006    \\ \hline
\multirow{2}{*}{1/3} & 5    & 2    & 8.9 $\pm$ 1.9             & \textbf{0.815 $\pm$ 0.005}     \\ \cline{2-5} 
                       & 10   & 1    & 9.6 $\pm$ 0.9           & \textbf{0.820 $\pm$ 0.006}      \\ \hline
\multirow{3}{*}{1}     & 5    & 6    & 8.9 $\pm$ 2.2           & \textbf{0.811 $\pm$ 0.005}      \\ \cline{2-5} 
                       & 10   & 3    & 9.9 $\pm$ 1.9           & \textbf{0.812 $\pm$ 0.005}      \\ \cline{2-5} 
                       & 30   & 1    & 13.4 $\pm$ 2.5          & 0.785 $\pm$ 0.006       \\ \hline
\multirow{3}{*}{2}     & 10   & 6    & 9.6 $\pm$ 1.6           & 0.799 $\pm$ 0.005     \\ \cline{2-5}
                       & 30   & 2    & 12.0 $\pm$ 2.3          & 0.791 $\pm$ 0.006      \\ \cline{2-5} 
                       & 60   & 1    & 16.8 $\pm$ 2.7          & 0.782 $\pm$ 0.006     \\ \hline
\end{tabular}

\caption{\label{tab:tab4}Overall results (AUC) on the validation simulator dataset obtained with the best models for various sequence hyperparameters ($L_k$, $S_k$) and $\Delta T_k = 1$ s. Significant results are highlighted in bold.}

\end{table}

The study shows that using a single input image yield unsatisfactory results ($AUC =$ 0.734).

Better results were achieved with a $1/3$-second window or a $1$-second window using $(L_k, S_k) = (10, 3)$ or $(L_k, S_k) = (5, 6)$, yielding significantly better results than video snippets with 30 frames $(L_k, S_k) = (30, 1)$ and reduced computation time by around 25 $\%$.

Tab.\ref{tab:tab5} provides results by surgical error classes with EfficientNetB0 + LSTM, 600$\times$600 pixel, $(L_k, S_k) = (10, 3)$ images for the same prediction horizon ($\Delta T_k = 1$ second) on validation and test data.

\begin{table*}[!]
\centering

\begin{tabular}{c|l|l|l|l|l|l|}
\cline{2-6}
\multicolumn{1}{l|}{}                & \multicolumn{1}{c|}{AUC (\textit{overall})} & \multicolumn{1}{c|}{AUC (\textit{smooth.})} & \multicolumn{1}{c|}{AUC (\textit{out of red reflex})} & \multicolumn{1}{c|}{AUC
(\textit{radial extension})} & \multicolumn{1}{c|}{AUC (\textit{tear})} \\ \hline

\multicolumn{1}{|c|}{\textit{validation}}  & 0.812 $\pm$ 0.005                          & 0.789 $\pm$ 0.006                           & 0.976 $\pm$ 0.003                                     & 0.698 $\pm$ 0.013                                      & 0.804 $\pm$ 0.026                       \\ \hline

\multicolumn{1}{|c|}{\textit{test}}  & 0.820 $\pm$ 0.005                           & 0.771 $\pm$ 0.006                            & 0.981 $\pm$ 0.003                                  & 0.682 $\pm$ 0.011                                   & 0.815 $\pm$ 0.025                      \\ \hline

\end{tabular}
\caption{\label{tab:tab5}Prediction results (AUC) for surgical errors by class, obtained with the best model on validation and test simulator data.}
\end{table*}

\begin{table*}[]

\centering
\resizebox{1\textwidth}{!}{
\begin{tabular}{ccc|l|l|l|l|}
\cline{4-7}
\multicolumn{1}{l}{}                                      & \multicolumn{1}{l}{}                                      & \multicolumn{1}{l|}{}                                                           & AUC (\textit{overall}) & AUC (\textit{smooth.}) & AUC (\textit{out of red reflex}) & AUC (\textit{radial extension}) \\ \hline
\multicolumn{2}{|c|}{\multirow{3}{*}{\textit{data homogenization}}}                                                            & \textit{central crop}                                                                    & 0.570         & 0.485         & 0.649                   & 0.430                  \\ \cline{3-7} 
\multicolumn{2}{|c|}{}                                                                                                & \textit{ocular border mask}                                                              & 0.615         & 0.559        & 0.678                   & 0.480                  \\ \cline{3-7} 
\multicolumn{2}{|c|}{}                                                                                                & \begin{tabular}[c]{@{}c@{}}\textit{ocular border mask} \\ + \textit{histo. matching}\end{tabular} & 0.561         & 0.502         & 0.639                   & 0.509                  \\ \hline \hline
\multicolumn{1}{|c|}{\multirow{7}{*}{\begin{tabular}[c]{@{}c@{}}\textit{features} \\ \textit{alignment}\end{tabular}}} & \multicolumn{1}{c|}{\multirow{4}{*}{\begin{tabular}[c]{@{}c@{}}\textit{simulator} \\ \textit{(source)}\end{tabular}
}}  & $\emptyset$                                                                              & \textbf{0.784}         & \textbf{0.715}         & 0.950                   & \textbf{0.647}                  \\ \cline{3-7} 
\multicolumn{1}{|c|}{}                                    & \multicolumn{1}{c|}{}                                     & \textit{CORAL}                                                                           & 0.771         & 0.645         & \textbf{0.963}                   & 0.593                  \\ \cline{3-7} 
\multicolumn{1}{|c|}{}                                    & \multicolumn{1}{c|}{}                                     & \textit{MMD}                                                                             & 0.747         & 0.700         & \textbf{0.967}                   & 0.602                  \\ \cline{3-7} 
\multicolumn{1}{|c|}{}                                    & \multicolumn{1}{c|}{}                                     & \textit{MSE}                                                                             & 0.765         & \textbf{0.719}         & 0.956                   & 0.617                  \\\cline{2-7} \cline{2-7} \cline{2-7} 
\multicolumn{1}{|c|}{} 
& \multicolumn{1}{c|}{\multirow{3}{*}{\begin{tabular}[c]{@{}c@{}}\textit{real} \\ \textit{surgery} \\ \textit{(target)}\end{tabular}
}} & \textit{CORAL}                                                                           & 0.654         & 0.611         & 0.710                  & 0.751                 \\ \cline{3-7} 
\multicolumn{1}{|c|}{}                                    & \multicolumn{1}{c|}{}                                     & \textit{MMD}                                                                             & 0.651         & 0.605         & 0.685                   & 0.629                  \\ \cline{3-7} 
\multicolumn{1}{|c|}{}                                    & \multicolumn{1}{c|}{}                                     & \textit{MSE}                                                                             & 0.542         & 0.517         & 0.645                   & 0.518                  \\ \hline
\end{tabular}}
\caption{\label{tab:tab6}AUC results (domain adaptation) for data homogenization and feature alignment. Significant results are highlighted in bold.}
\end{table*}

\begin{figure}[!]
\centering
\includegraphics[width=0.49\textwidth]{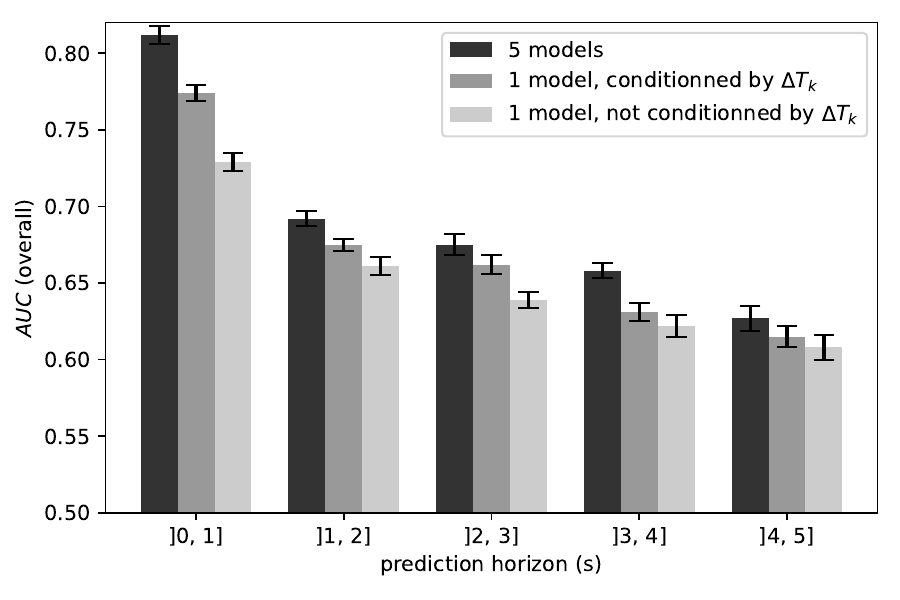}
\caption{\label{fig:fig7}Validation results (AUC) on the simulator dataset, obtained with the best configuration (algorithm and hyperparameters) for various prediction horizons.
}
\end{figure}

Fig.\ref{fig:fig7} shows the performance across different prediction horizons $\Delta T_k \in \llbracket 1,5 \rrbracket$ for 3 different training strategies.

The first strategy involves training one model for each prediction horizon. The second strategy consists of training a single model sequentially, starting with $\Delta T_k = 1$ and incrementally progressing up to $\Delta T_k = 5$. The third strategy entails this sequential training without providing the temporal horizon as input to the model during the training process.

\subsection{Surgical error prediction transfer to real surgical videos}

Due to limitations in computation time and memory capacity, we consider Inception$\_$ResNetv2 and use images of size 299$\times$299 pixels for unsupervised domain adaptation. 
We set $L_k = 10$, $S_k : 3$ and $\Delta T_K = 1$ second.

Tab.\ref{tab:tab6} presents the unsupervised domain adaptation results, using data homogenization and feature alignment strategies on validation dataset. The prediction horizon is fixed at 1 second for inference. We provide the overall results and those for the three specific surgical error classes observed in real surgical videos.

The \textit{central crop} of the real surgical images in video snippets to match the simulator's format served as the baseline, representing the lower performance bound for transfer.

Adding an \textit{ocular border mask} improved performance across all surgical error classes (+6.5$\%$ \textit{overall}). 

The \textit{histogram matching} strategy does not seem to have provided any benefit, except for the \textit{radial extension} class (+18$\%$), but the AUC of 0.509 indicates a performance only slightly better than random.

Regarding feature alignment, the row labeled \textit{source} corresponds to the validation results for the simulator data, while the row labeled \textit{real surgery target} indicates the results on the real surgery validation set. Note that the symbol $\emptyset$ represents the results obtained without using the discrepancy loss.

Tab.\ref{tab:tab7} presents results for feature alignment using the CORAL discrepancy loss after pre-taining the model with 2 different pretext tasks on validation and test data.

\begin{table*}[!]
\centering
\begin{tabular}{c|l|l|l|l|}
\cline{2-5}
\multicolumn{1}{l|}{}                                & \multicolumn{1}{c|}{AUC (\textit{overall})} & \multicolumn{1}{c|}{AUC (\textit{smooth.})} & \multicolumn{1}{c|}{AUC (\textit{out of red reflex})} & \multicolumn{1}{c|}{AUC (\textit{radial extension})} \\ \hline
\multicolumn{1}{|c|}{\textit{Frame order}}                    & 0.663                               & 0.632                               & 0.757                                         & 0.658                                        \\ \hline

\multicolumn{1}{|c|}{\textit{Image reconstruction}} & 0.657                                  & 0.617                                  & 0.774                                            & 0.674                                          \\ \hline
\end{tabular}
\caption{\label{tab:tab7}AUC results for feature alignment using CORAL discrepancy loss with pre-training on the validation real surgery dataset.}

\end{table*}

\section{Discussion}

This study aimed to predict real-time surgical errors from monocular cataract surgical videos by leveraging knowledge transferred from simulator data, with only a limited amount of labeled data from actual surgical videos. The use of a simulator offers a wide range of scenarios, allowing for more comprehensive predictions.

As expected, prediction becomes more challenging as the prediction horizon increases (e.g., $AUC = 0.812$ for $\Delta T_k = 1$ second versus $0.692$ for $\Delta T_k = 2$ seconds). The use of a time-step model yields significantly better results for prediction horizons $\Delta T_k$ ranging from 1 to 5 seconds. However, it has been shown that if a single model is employed, conditional predictions result in reduced performance decline for the shorter prediction horizons ($\Delta T_K$ = 1, 2, and 3 seconds). Our results suggest that having an extensive temporal history is unnecessary for accurate surgical error prediction. A short history of just 1 second proved to be sufficient, which is advantageous for real-time analysis from a computational point of view. Moreover, downsampling using every second frame did not degrade performance, enabling us to process video snippets with fewer frames and thus accelerating computations, which is beneficial for real-time analysis. However, an analysis based solely on individual images seems insufficient, incorporating at least a basic level of temporal dynamics is essential. According to our comparative study, using 600$\times$600 pixel images resulted in significantly better overall results. The best performance was achieved with EfficientNetB0, pre-trained on noisy student as spatial encoder, and an LSTM as temporal encoder. This configuration yielded an AUC score of 0.820 for a 1-second prediction horizon on the test set with $L_k = 10$ and $S_k = 3$. Regarding surgical error classes, we found that not all surgical errors are equally predictable. For instance, the \textit{out of red reflex} error is easier to predict as it depends on the eye’s orientation, i.e. general geometric features the model can readily capture.

Regarding the transfer to real surgery videos, the results on real surgical videos are promising, validating this initial attempt of knowledge transfer for the prediction task. We observe that the homogenization strategy, specifically adding an ocular border mask, improves performance compared to the baseline (AUC = 0.615 vs. 0.578 overall). This mask enhances the structural consistency of the images by adding a black pixel border. Simple histogram matching, however, does not appear to be effective, yielding lower performance than the ocular border mask alone ($AUC =$ 0.561 vs. 0.578 overall). This may be due to the overly strict constraint of histogram matching and the global, rather than local, changes in light intensity. The differences between the domains are not only explained by luminescence but also by the structure's shape, which is not altered by functions that act on the histogram. We also observe that aligning latent representations using CORAL and MMD yields better performance than data homogenization alone (AUC =  0.654 and 0.651 overall vs 0.615). The results obtained with these discrepancy losses are better than those with MME, which does not allow unsupervised domain adaptation in this work. It is important to note the slight drop in performance on the source domain compared to results without latent representation alignment (AUC = 0.771 with CORAL, 0.747 with MMD vs 0.784 overall). This decline is expected, as aligning latent representations involves a trade-off between optimizing surgical error prediction and regularizing the latent space. However, the degradation is minimal, highlighting that the transfer does not come at the cost of accurate surgical error prediction. Surprisingly, better performance (+1.5$\%$) was achieved for the out-of-red reflex error when applying domain adaptation compared to when no adaptation ($\emptyset$) was performed. This is likely due to the training strategy—using five models for transfer learning and retaining the best results, compared to using one without transfer—rather than the effects of the discrepancy loss. The temporal pretext task followed by domain adaptation with the CORAL loss improves overall performance (+2$\%$) and for \textit{smoothness} (around $+3 \%$), but especially for the \textit{out of red reflex} error, with an improvement of nearly 7$\%$. On the other hand, the image reconstruction-based pretext task shows an advantage mainly for the \textit{out of red reflex} error, with a 9$\%$ increase. This is not surprising, as this error class is easier to transfer because it relies primarily on geometric features, which are more straightforward to generalize across domains.


While our approach focused on data alignment, future research could explore alternative strategies such as data augmentation. Data augmentation could help addressing both inter-domain and intra-domain variability, potentially leading to further performance improvements. Finally, future research could explore integrating advanced information beyond RGB channels, such as detailed information about the capsulorhexis process itself (identifying the capsulorhexis opening). Additionally, the use of binocular video could further enhance surgical error prediction.  Binocular videos provide richer data by capturing depth and spatial relationships more effectively, which can facilitate more accurate analysis of surgical movements and errors. This additional perspective could give a more comprehensive view of the surgical field, potentially enhancing model performance and the reliability of predictions. 

A major limitation is the very low number of errors in real surgeries, which makes validation and testing challenging, and comparison to a supervised domain adaptation approach impossible. Moreover, although the simulator contains many errors compared to a real surgery dataset, the classes are highly imbalanced. Novel strategies should be established to address this challenge.

\section*{Compliance with ethical standards}

This study was conducted following the Declaration of Helsinki. Consent was obtained from participants to allow this study to take place.

\section*{Acknowledgments}

The authors thank Haag-Streit for making this work possible by sharing the simulator data and authorizing it. We also wish to express our gratitude to the ophthalmology department of the University Hospital of Brest for their collaboration and for providing essential data for our study. This work was supported by Inserm and the Brittany Region through the ARED program (ASCIA project).


\bibliographystyle{elsarticle-num-names} 
\bibliography{cas-refs}

\end{document}